\documentclass[twocolumn,nofootinbib,aps,showpacs,showkeys]{revtex4}

\usepackage{graphicx}
\usepackage{dcolumn}
\usepackage{bm}
\usepackage{graphics}
\usepackage{amsfonts}

\begin{document}

\preprint{}

\title{Renormalization-group running of the cosmological constant 
and the fate of the universe}

\author{B. Guberina}
\email{guberina@thphys.irb.hr}
\author{R. Horvat}
\email{horvat@lei3.irb.hr}
\author{H. \v Stefan\v ci\' c}
\email{shrvoje@thphys.irb.hr}
\affiliation{%
Rudjer Bo\v skovi\' c Institute, P. O. Box 180, HR-10002 Zagreb, Croatia\\
}%


\date{\today}

\begin{abstract}
For a generic quantum field theory we study the role played by the
renormalization-group (RG) running of the cosmological constant (CC)
in determining the ultimate fate of the universe. We consider
the running of the CC of generic origin ( the vacuum energy of quantum fields 
and the potential energy of classical fields ), 
with the  
RG scale proportional to the (total energy density$\rm{)^{1/4}}$ as the most
obvious identification. Starting from the present-era values for
cosmological parameters we demonstrate how the running can easily provide a
negative cosmological constant, thereby changing the fate of the universe,
at the same time rendering compatibility with critical string theory. 
We also briefly discuss the recent past in our scenario.
\end{abstract}




\maketitle



Recent Type Ia supernovae \cite{1} and the CMB data \cite{2} show 
that our universe is spatially flat $(\Omega_k = 0)$ and 
accelerating at present time. Specifically, the experimental 
situation at present is such that the present energy density 
consists of approximately 1/3 of ordinary matter and 2/3 of dark 
energy with a negative pressure. For this dark energy to accelerate
the expansion, its equation of state  $w \equiv p/\rho $ must
satisfy $w < - 1/3 $. 

The most promising candidate for this cosmic acceleration and 
dark energy seems to be a ``true'' cosmological constant (CC) 
$\Lambda $ (with $\Lambda  > 0 $), for which the equation of 
state is simply $w = -1 $. Another interpretation of dark
energy
relies on  the existence of an ultralight scalar field $\phi $
(``quintessence'') \cite{3}, in which case the dark energy density
is the result of the scalar field slowly evolving along its 
effective potential $V(\phi )$, with $w \approx -1 $. 
Some alternate explanations for the present
acceleration beyond dark energy include theories in more than four
dimensions \cite{4} as well as the Chaplygin gas \cite{5}. 

If the presently dominating dark energy and the acceleration of the universe  
are due to the ``true'' CC or ``quintessence'', the expansion will keep
accelerating perpetually, rapidly approaching a de Sitter (dS) regime.
This implies a fate for open $(\Omega_k > 0)$ and flat $(\Omega_k = 0)$ 
universes in which the scale factor expands exponentially, 
$a \sim e^{Ht}$, for an indefinitely long time. The same is destined even
for a closed universe, unless $\Omega_{\Lambda}$ is too low so that the 
universe collapses before the CC becomes predominant. More importantly, 
an eternal expansion and acceleration would exhibit future event 
horizons, which presents a rather serious challenge to critical string
theory \cite{6}. Namely, the finite event horizon implies the existence
of regions in space that are inaccessible to light probes, thereby 
preventing the definition of asymptotic states and therefore the construction of 
the conventional $S$-matrix description required for field theory. This 
implies the impossibility of formulating the $S$-matrix for strings, which are
by definition theories of the on-shell $S$-matrix, in such backgrounds. A 
noncritical string theory, characterized by a ``graceful exit'' from the
inflationary phase, has recently been considered in \cite{7}.
 
Nonetheless, there are other models describing dark energy in which the dS
phase is transient, thereby bringing back compatibility with string 
theory. One loophole providing an alternative to a perpetually 
accelerating universe is to have a potential $V(\phi )$ which has a 
minimum at $\phi_0 $, with $V(\phi_0 ) = 0 $ and $V^{''}(\phi_0 ) > 0 $.  
In this case, during the coherent oscillation phase, the energy density
decreases as $a^{-3}$ $(w = 0 )$, leading to a 
decelerating universe. In another scenario, $V(\phi )$ has a minimum 
$V(\phi_0 )$ = 0 and then becomes flat for $\phi  > \phi_0 $. After 
passing the minimum, the epoch of kinetic energy domination sets in 
where the energy density scales as $a^{-6}$ $(w = 1 )$, resulting
again in a decelerating universe. Nonetheless, the simplest case is to 
have a negative CC. In this case, no matter the universe is open, flat, 
or closed, or how extremely tiny the CC is,  
it collapses eventually, thereby bypassing the troublesome 
problems with the event horizon. As a way of example, we mention 
a mechanism to 
generate a negative CC in \cite{8} (for supergravity theories, see, 
e.g., \cite{9}).

In the present paper, we expand our previous work \cite{BGHS} on the RG running
of the CC in the standard model, to a generic field theory model. We then
study the future of the universe in such a generic scenario. Recent works
\cite{ShSoPLB, BGHS, JHEP} 
show that even a ``true'' CC cannot be fixed to any definite
constant (including zero), owing to RG running effects. Particle contributions
to the RG running of $\Lambda $ due to vacuum fluctuations of massive fields 
have been properly derived in \cite{BGHS}, with
a somewhat unexpected outcome that more massive fields do play a dominant
role in the running at any scale. If our chosen value for the RG scale $\mu
$ is always below the lowest mass in the underlying quantum field  theory, 
then the scaling evolution
of the CC $\Lambda (\mu )$ towards $\mu \rightarrow  0 $
is essentially given only by two parameters which depend on the
field content (particle masses as well as the total number of massive degrees of
freedom including the +/- sign for bosons/fermions). The same structure of the
scaling behavior is also expected for $\Lambda $ arising from vacuum
condensates of scalar fields \cite{BGHS}, in which case there is an additional
dependence of the parameters on the particle interaction. We show that even if
we start with a positive $\Lambda $ in a flat universe, 
the scaling behavior may naturally be
such as to provide a negative $\Lambda $ for an indefinitely long time,
thereby changing the ultimate fate of the universe.

We have already mentioned the  possibility that a flat universe may collapse in the
future. In most models of dark energy based on extended supergravity, this
occurs in the nearby future, within the next few billion years \cite{10}.
Although provided with a negative CC for an indefinitely long time, our
scenario allows quite a distinctive feature: the universe never
collapses, starting to decelerate relatively soon and increasing the scale
factor to its maximum value in infinite time.

      
In the formalism of quantum field theory in curved space-time, the cosmological
constant receives divergent contributions originating from {\em zero-point
energies} of the quantum fields considered. Being one of the parameters in 
the Lagrangian, the cosmological constant undergoes renormalization
(for earlier works on the subject see e.g. \cite{odin}). After the
appropriate choice of the renormalization scheme which correctly comprises the
effects of mass thresholds \cite{BGHS}, 
one can formulate the renormalization-group equation in the form

\begin{equation}
\label{eq:runneq}
(4 \pi)^2 \mu \frac{\partial \rho_{\Lambda}}{\partial \mu} = 
\sum_{i} \frac{I_{i}}{2} m_{i}^{4} \frac{\mu^2}{\mu^2 + m_{i}^2} \, ,
\end{equation}
where $\mu$ is the renormalization scale, $\rho_{\Lambda}$ represents the energy
density attributed to the cosmological constant, and $i$ is an index of any
massive degree of freedom (corresponding to the quantum field), 
both fermionic and
bosonic. The quantity $I_{i}$ acquires the value $+1$ for bosonic and $-1$ for
fermionic degrees of freedom. It is clear that more massive fields dominate
(\ref{eq:runneq}) at any scale and this departure from the Appelquist-Carazzone
theorem \cite{AppCarr} results from the $(mass)^4$ dimensionality of 
$\rho_{\Lambda}$ \cite{BGHS}. The following comment is in order. It is the
high dimensionality of the CC energy density that causes the failure of the decoupling theorem;
the peculiar mass-dependent renormalization scheme we used in \cite{BGHS} in
order to renormalize the zero-point energy agrees with the $\rm \overline{MS}$
scheme in the high-energy limit. The result is furthermore supported by the
failure of the decoupling theorem in the case of vacuum condensate of the Higgs
field \cite{BGHS} where the mass-dependent renormalization scheme is unambiguous
\footnote{Recently, Gorbar and Shapiro \cite{GorShap} studied the form of 
decoupling
for the CC energy density in quantum field theory in the external
gravitational field on a flat
background. The results fail to reproduce a correct high-energy behavior as
given in the $\rm \overline{MS}$ scheme, and, therefore, cannot be conclusive.}.

In the approach described above, the particle zero-point energies 
 are taken as a generic contribution to the running of the cosmological
 constant. The results for the 
 running are obtained at the one-loop level. In the framework we used, 
 quantum field theory of matter (particles) is defined on the curved
 classical background which is described through the vacuum action
 \begin{eqnarray}
 \label{eq:vacac}
 S & = & \frac{1}{16 \pi G} \int d^4 x \sqrt{-g} (R - 2\Lambda ) + 
 \nonumber \\
   & & \int d^4 x \sqrt{-g} \left[ a_1 R_{\mu\nu\rho\sigma}^2 
   + a_2 R_{\mu\nu}^2 +a_3 R^2 + a_4 \Box R \right] \, .
 \end{eqnarray}

 This equation, however, does not include the possible contributions of 
 quantum gravity. As discussed in \cite{JHEP}, one could safely neglect 
 the gravity  effects
 related to   all terms in (\ref{eq:vacac}) proportional to $a_i$,  
 and the nonminimal term
 $\xi \Phi^\dagger \Phi R$ {\it at the present time and scale}. 
However, nonperturbative effects of quantum gravity  
 could play a certain role even today.
 
 Althogh it is presently well known that there are tremendous difficulties
 in the formulation of quantum field theory of gravity, it is still possible 
 to renormalize the nonrenormalizable theories in the framework of 
 effective field theories \cite{Wein}. A successful 
 example of such a theory is a chiral perturbation theory as a low-energy 
 nonlinear realization of QCD. Gravity is potentially even a better 
 effective theory, since quantum corrections appear to be rather small,
 persisting so up to the Planck scale. It is then possible, using the effective
 field theory to calculate the 
 effects of quantum gravity which emerge from the low-energy 
 part of the theory without knowing the "true" quantum field theory of
 gravity \cite{Don}.
 
 What really makes the effective field theory of gravity different from  
 other effective theories is the eventual existence of singularity 
 \cite{Hawk} in the 
 future - which, in the limit of extremely low energy (corresponding to the the 
 wavelength probed of the order of the size of the universe), breaks the 
 validity of the theory in the infrared limit. For example, conformal 
 invariance breaking (which is also broken by gravitons!) causes infrared
 instability in de Sitter space \cite{Einhorn}.  
 Therefore quantum gravity may lead to strong renormalization effects at
 large distances owing to the infrared divergences \cite{Tsamis}.

 Recently, Bonanno and Reuter \cite{Bon} derived RG equations with an 
 infrared cutoff which at a given scale stops the running 
 of $\Lambda$ in the
 infrared. The basic framework is the effective average action  
 $\Gamma_k [g_{\mu\nu} ]$, which is  a Wilsonian coarse grained free energy.
 It depends on a momentum scale $k$, and defines an effective 
 field theory appropriate for the scale $k$. An additional hypothesis
 in the formulation of the RG behavior of $\Lambda$ and $G$ is that, in the 
 limit $k \rightarrow 0$, both $\Lambda (k)/k^2$ and $k^2 G(k)$ run into 
 an infrared attractive non-Gaussian fixed point.

 The scale $k$ plays a role of an infrared cutoff \cite{Bon}. 
 Intuitively, an infrared
 cutoff is necessary since, for example, at the present time $t=t_0$ ($t=0$ at 
 the Big Bang), one has to integrate quantum fluctuations  down to the scale 
 $k > \frac{1}{t_0}$; larger scales for which $k < \frac{1}{t_0}$ should 
 obviously be truncated.

 However, it is very difficult to imagine the precise physical mechanism which 
 acts as a cutoff. It would depend on 
 the problem under consideration and may include many scales including, e. g., 
 particle momenta, field strengths, the curvature of spacetime, etc.
 We address this problem and our choice of the scale in the discussion later on.

The integration of Eq. (\ref{eq:runneq}) gives a closed form of the cosmological
constant energy density at any value of the renormalization scale $\mu$:

\begin{equation}
\label{eq:lamatmu}
\rho_{\Lambda}(\mu) - \rho_{\Lambda}(0) = \frac{1}{4 (4\pi)^2} \sum_{i} I_{i}
m_{i}^{4} \ln \left( 1 + \frac{\mu^2}{m_{i}^2} \right) \, .
\end{equation}
At renormalization scales much smaller than the mass of the lightest quantum
field the expansion in powers of $\mu^2$ gives the following form for 
$\rho_{\Lambda}$:

\begin{eqnarray}
\label{eq:runnatsmallmu}
\rho_{\Lambda}(\mu) - \rho_{\Lambda}(0) & = & \frac{1}{4 (4\pi)^2} \left[
\left( \sum_{i}^{N_{b}} m_{i}^{2} - \sum_{j}^{N_{f}} m_{j}^{2}
\right) \mu^{2} \right. \nonumber \\
& - & \left.  \frac{1}{2} (N_{b} - N_{f}) \mu^{4} + {\cal O}\left(
\frac{\mu^{6}}{m_{lightest}^2} \right) \right] \, .
\end{eqnarray}
In the equation given above, $N_{b}$ and $N_{f}$ denote the 
total numbers of bosonic and
fermionic massive degrees of freedom, respectively.

The problem to be solved next  is the choice of the renormalization 
scale $\mu $ in cosmological settings. In an evolving universe, it would 
be natural to identify the renormalization scale $\mu $, representing typical 
particle (graviton) momenta in the universe, with some energy scale 
characteristic of the stage of  evolution of the universe. During epochs 
of radiation domination, association of the scale $\mu $ with temperature, 
$\mu \sim T $, is by all means straightforward. In our approach, we adopt 
as a natural choice
\begin{equation}
\mu = \rho^{1/4} \, ,
\label{eq:defmu}
\end{equation}
where $\rho $ stands for the total energy density of the universe. This choice
determines the renormalization scale in terms of a  quantity well defined in 
cosmological settings. Using (\ref{eq:defmu}) we obtain a consistent definition 
of the 
scale $\mu$ for all epochs of the evolution of the universe. Moreover, in the
radiation dominated universe, the definition of the renormalization scale 
(\ref{eq:defmu}) coincides with the straightforward association $\mu \sim T$ 
mentioned above.  

Next we discuss our choice of the scale, $\mu = \rho^{1/4}$, and compare it
with other possible choices, especially with that proposed by Shapiro and
Sola \cite{JHEP}, $\mu \sim \sqrt{R} \sim H$, where $R$ is the curvature
scalar and $H$ denotes the Hubble parameter.

Let us recall that in the decoupling theorem, it is the momentum of 
particles entering a quantum loop, and the mass of  particles inside the
loop, that are to be compared (with massless gravitons on external legs in
the case of CC). On the other hand, the coupling constants play no role 
in the decoupling. It has been argued \cite{JHEP} that the value of the
curvature scalar $R$ can be taken as an order-parameter for the gravitational
energy, i.e., $\mu \sim \sqrt{R}$. The Einstein equations imply that, at
present,
\begin{equation}
\mu_0 \sim R_0^{1/2} \sim (G \rho_0 )^{1/2} \sim H_0 \;,
\label{eq:shapiro}
\end{equation}
where $G$ denotes the Newton's constant. According to the choice 
(\ref{eq:defmu}), the relevant scale cannot be less than  that of
typical momenta of background gravitons, which, at present, are of the order of
$10^{-4} \;\mbox{\rm eV}$ and not of $H_0  \simeq 10^{-33} \;\mbox{\rm eV}$.
We notice that the large separation of the scales (\ref{eq:defmu}) and 
(\ref{eq:shapiro}) is actually due to
a coupling of gravitons to ordinary matter ($\sim G E^2 $ with $E$ being a
typical graviton energy), inherently present in the expression (\ref{eq:shapiro}). 
Indeed, one finds 
that the ratio of scales (\ref{eq:shapiro}) and (\ref{eq:defmu}) 
is essentially given by the square root of the
graviton coupling constant. It is thus pretty obvious that the Shapiro-Sola 
choice (5) means a restriction to  ``soft'' graviton momenta, whereas our
choice (4) involves typical (or ``hard'') graviton momenta. \footnote{The
above separation of the scales resembles a distinction between hard ($\sim T$)
and soft ($\sim gT$) momenta encountered in quantum field theories at finite
temperature/density, where $g$ is the gauge coupling constant of the
underlying theory \cite{11}.} In turn, this implies that the choice
(\ref{eq:shapiro})
corresponds to CC's that are nearly uniform in both space and time, 
thus being compatible with models which treat $\Lambda $ 
as an (almost) true
constant. In this case, $\Lambda $ acts as a source of gravitons with
extremely large wavelengths, of the order $H_{0}^{-1}$. 
Moreover, with the choice (\ref{eq:shapiro}), the running of the CC on cosmological scales is
essentially negligible. On the other hand, adopting the choice (\ref{eq:defmu}),
one arrives at a model with the  
pronounced running of $\Lambda $, comparable with that of 
ordinary matter. 
This feature is clearly visible in our figures, see below.   

Let us in more detail discuss our advocacy for the scale as given by
(\ref{eq:defmu}). First of all, we would like to have the correct limit towards the
past radiation epoch, where all relevant particle scales are essentially
given in terms of the temperature of interacting particle species. Obviously,
this is not possible with the choice (\ref{eq:shapiro}). In addition, in our approach
the energy-momentum tensor $T_{\mu \nu }$ is no longer separately
covariantly conserved, but only is the sum $T_{\mu \nu } + \frac{\Lambda }{8
\pi G} g_{\mu \nu }$ (see  (\ref{eq:ZSEdetailed}) below). 
This implies that there exists an
interaction between  matter and the CC which causes a continuous transfer of
energy from matter to the CC or vice versa, depending on the sign of the
interaction term. We shall see below that in our approach the CC and matter are
always tightly coupled, with a consequence that the CC actually behaves more like
ordinary matter than a ``true'' CC. Consequently, both components are
assumed to act as short wavelength sources of gravitons with $\mu >>
H \sim 1/t $ ($H \sim 1/t $
is adequate only for the CC with a negligible interaction with ordinary
matter). A similar distinction of scales can be found in the paper \cite{Bon},
with the statement that the choice $\mu \sim H \sim 1/t $ is adequate only
for the case of perfect homogeneity and isotropy, otherwise new scales much
larger than the cosmological one need be introduced. Finally, the choice $\mu
\sim H \sim 1/t $, in our approach, which
aims to take only effects of matter on the renormalization of $\Lambda $,
causes a few problems on their own. For instance, if one identifies the
parameter $A$ ((\ref{eq:runnatsmallmu}) and (\ref{eq:lambdamu}) below) 
with the square of the highest mass in the underlying
particle theory, one can fit the present experimental data for
$\rho_{\Lambda}$, $\rho_{\Lambda}(\mu = \mu_0 )\sim 10^{-12}
\;\mbox{\rm eV}^4 $,
only with $A \sim M_{Pl}^2 $, which is by no means plausible(obviously, the
Planck scale is natural in the context of quantum gravity but not in the
context of any particle theory). The only remedy would be to give up the
``nice'' infrared limit, $\rho_{\Lambda}(0) = 0$, which necessitates antropic
considerations to explain the nonzero infrared value.
Furthermore, as already stated, in this case one finds that
$\rho (\mu = \mu_0 ) \simeq \rho_{\Lambda}(0)$, which leads to totally
negligible running, therefore rendering this case completely uninteresting
for
the purpose of the paper, i.e., for the study of the fate of the universe.

Clearly, the choice (\ref{eq:defmu}) constitutes a viable model as long as 
the total energy density is
positive. Therefore, we consider the evolution of the universe in a time
interval in which the identification of Eq. (\ref{eq:defmu}) is sensible.
As we shall see below, the corresponding time interval turns out to be the whole
interval $t_{0} < t < \infty$, and therefore the choice of the renormalization
scale given by (\ref{eq:defmu}) actually poses no restriction at all.

From this point on we consider the evolution of the universe in cosmology with
the running (indirectly time-dependent) cosmological constant energy density. We
assume that the present value of the cosmological scale 
$\mu_{0} \simeq 2 \times 10^{-3} \, {\rm eV}$ is
sufficiently smaller than the lightest of the masses of the fields in the theory.
In this case, the asymptotic expansion (\ref{eq:runnatsmallmu}) applies.
In view of recent observations of neutrino oscillations by Superkamiokande
\cite{12} and SNO \cite{13}, we find the expansion (\ref{eq:runnatsmallmu})
justifiable as these experiments seem to indicate that neutrinos do have a 
significant mass above $10^{-3} \, {\rm eV}$.
In addition, we assume that the infrared limit of the 
cosmological constant energy density
vanishes, $\rho_{\Lambda} (0) = 0$, hence avoiding anthropic considerations 
to explain this value. In this way, we arrive at a model in which the
cosmological constant energy density is dynamically determined at any scale.
Although the coefficient of the $\mu^2 $-term in (3) is severely restricted
by the big bang nucleosynthesis results \cite{BGHS}, this term, if nonzero,
still dominates at sufficiently low scales. Therefore, at sufficiently low  
scales $(\mu_0
\geq \mu \geq 0)$, relevant to the study of the ultimate fate of the universe
(i.e., for the corresponding times $t_0 \leq t \leq \infty)$, we describe the 
running of the cosmological constant energy density by the first two terms 
in the expansion (\ref{eq:runnatsmallmu})

\begin{equation}
\label{eq:lambdamu}
\rho_{\Lambda} = A \mu^{2} + B \mu^{4} \, ,
\end{equation}
which together with the definition of the cosmological energy scale 
(\ref{eq:defmu}) gives

\begin{equation}
\label{eq:lambdarho}
\rho_{\Lambda} = A \rho^{1/2} + B \rho \, .
\end{equation}
We have shown in the previous paper \cite{BGHS} that also for the
vacuum-condensate contribution to $\Lambda $ one should expect the same
scaling behavior of $\rho_{\Lambda }$ as in (\ref{eq:lambdamu}). 
Hence, one may consider (\ref{eq:lambdamu})   
a scaling behavior for generic $\Lambda $, where, besides the definition in
(\ref{eq:runnatsmallmu}), 
the coefficients $A$ and $B$ now take on an extra dependence on the particle 
interaction.
In order to round up the survey of our cosmological model there remains to
identify other components of the total energy density apart from the
cosmological constant energy density. In a general manner, we express the total
energy density and pressure of the content of the universe in a form

\begin{eqnarray}
\label{eq:rhocomponents}
\rho & = & \rho_{\Lambda} + \rho_{m} \, , \nonumber \\
p & = & p_{\Lambda} + p_{m} \, ,  
\end{eqnarray}
where we have used the symbols $\rho_{m}$ and $p_{m}$ to specify the density and
pressure of all other components of the universe, apart from the dark energy
ones. 
Thus we assume that the running cosmological constant is the only source
of the present acceleration of the universe, i.e., the only dark energy
component. 

In order to couple the cosmological term (being a dynamical quantity) with the
Einstein field equations
\begin{equation}
\label{eq:genrel}
G_{\mu\nu} + \Lambda g_{\mu \nu} = - 8\pi G T_{\mu \nu} \, ,
\end{equation}
we completely adopt the prescription outlined in a review paper 
\cite{overduin} to
move $\Lambda $ to the right-hand side of (\ref{eq:genrel}), interpreting it, 
at the same
time, as a part of the matter content of the universe, rather than a pure
geometrical entity. After doing so \footnote{If one rather prefers to keep a
variable $\Lambda $-term on the left-hand side of (\ref{eq:genrel}), 
then introduction of
some additional terms into (\ref{eq:genrel}) is necessary, which 
are functions of the scalar in a tensor-scalar theory.}, 
it is the effective energy-momentum
tensor $T_{\mu \nu} + \frac{\Lambda}{8 \pi G} g_{\mu \nu}$ 
that has the perfect fluid form and therefore  
satisfies energy conservation:
\begin{equation}
\label{eq:ZSEdetailed}
a \frac{d \rho}{d a} + 3 (\rho_{m} + p_{m}) = 0 \, .                                      
\end{equation}                                        
The equation of state obeyed by the cosmological term is then of the
familiar form, $p_{\Lambda } = - \rho_{\Lambda }$, and that obeyed by
ordinary matter may have a general form $p_m = w \rho_m $ (
although in our analysis we restrict ourselves to the pressure-free Friedmann
models with $w = 0$).
Note from (\ref{eq:ZSEdetailed}) that $\rho_m $ scales in a 
familiar way, $\rho_m \sim a^{-3}$,
when $\Lambda = constant $ as well as $w = 0$.

Furthermore, we introduce a condition (at $a(0) = a_{0}$) coming from
present observational results on the share of dark energy in the total 
energy density at the present epoch:

\begin{equation}
\label{eq:initial}
\rho_{\Lambda} (a = a_{0}) = \Omega_{\Lambda}^{0} \rho_{0} \, ,
\end{equation}
with $\Omega_{\Lambda}^{0} \simeq 0.7$. The condition given above determines the
parameter $A$ in (\ref{eq:lambdarho})

\begin{equation}
\label{eq:A}
A = (\Omega_{\Lambda}^{0} - B) \rho_{0}^{1/2} \, .
\end{equation}
With substitutions $u_{\rho} = \rho/\rho_{0}$ and $y = \ln (a/a_{0})$, 
Eq. (\ref{eq:ZSEdetailed}) acquires the form

\begin{equation}
\label{eq:master}
\frac{d u_{\rho}}{d y} + 3(1+w)((1-B)u_{\rho} - (\Omega_{\Lambda}^{0} - B)
u_{\rho}^{1/2}) = 0 \, ,
\end{equation}
while the initial condition is $u_{\rho} (y=0) = 1$. This equation has the
following solution:

\begin{equation}
\label{eq:rhoversusa}
u_{\rho} = \frac{\rho}{\rho_{0}} = 
\left[ \frac{\Omega_{\Lambda}^{0} - B}{1 - B} +
\frac{1-\Omega_{\Lambda}^{0}}{1 - B} \left( \frac{a}{a_{0}}
\right)^{-\frac{3}{2}(1-B)(1+w)} \right]^{2} \, .
\end{equation}
The expressions for energy densities $\rho_{\Lambda}$ and $\rho_{m}$ are

\begin{eqnarray}
\label{eq:loda}
\rho_{\Lambda} & = & \rho_{0} \left[ \left(\frac{\Omega_{\Lambda}^{0}-B}{1-B} 
\right)^2 \right. \nonumber \\ 
& + & \frac{(1+B)(\Omega_{\Lambda}^{0} -
B)(1-\Omega_{\Lambda}^{0})}{(1-B)^{2}} \left(\frac{a}{a_{0}}
\right)^{-\frac{3}{2} (1-B) (1+w)}  \nonumber \\ 
& + & \left. B \left(\frac{1-\Omega_{\Lambda}^{0}}{1-B} \right)^{2} \left(
\frac{a}{a_{0}} \right)^{-3 (1-B)(1+w)} \right] \, ,
\end{eqnarray} 
\begin{eqnarray}
\label{eq:moda}
\rho_{m} & = & \rho_{0} \left[   \frac{(\Omega_{\Lambda}^{0} -
B)(1-\Omega_{\Lambda}^{0})}{(1-B)} \left(\frac{a}{a_{0}}
\right)^{-\frac{3}{2} (1-B) (1+w)} \right. \nonumber \\ 
& + & \left. \frac{(1-\Omega_{\Lambda}^{0})^{2}}{1-B}  \left(
\frac{a}{a_{0}} \right)^{-3 (1-B)(1+w)} \right] \, .
\end{eqnarray}
In the special case $B=1$, the solution of (\ref{eq:master}) has the form

\begin{equation}
\label{eq:rhoB1}
u_{\rho} = \frac{\rho}{\rho_{0}} = \left[ 1 - \frac{3}{2}
(1-\Omega_{\Lambda}^{0})
(1+w) \ln \left( \frac{a}{a_{0}} \right) \right]^{2} \, .
\end{equation}
The expressions for energy densities $\rho_{\Lambda}$ and $\rho_{m}$ in the case
$B=1$ are

\begin{eqnarray}
\label{eq:lodaB1}
\rho_{\Lambda} & = & \rho_{0} \left[ \Omega_{\Lambda}^{0} - \frac{3}{2}
(1+\Omega_{\Lambda}^{0})(1-\Omega_{\Lambda}^{0})(1+w) \ln \left( \frac{a}{a_{0}}
\right) \right. \nonumber \\
& + & \left. \frac{9}{4} (1-\Omega_{\Lambda}^{0})^{2} (1+w)^{2} \left( \ln
\left(\frac{a}{a_{0}} \right) \right)^{2} \right] \, ,
\end{eqnarray}
\begin{equation}
\label{eq:modaB1}
\rho_{m} = \rho_{0} (1-\Omega_{\Lambda}^{0}) \left[ 1 - \frac{3}{2}
(1-\Omega_{\Lambda}^{0})(1+w) \ln \left( \frac{a}{a_{0}} \right) \right] \, .
\end{equation}  
\begin{figure}
\centerline{\rotatebox{-90}{\resizebox{0.4\textwidth}{!}
{\includegraphics{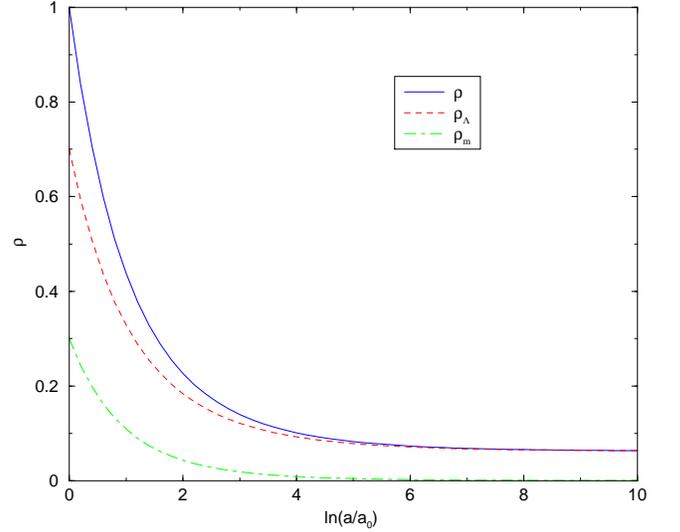}}}}
\caption{\label{fig:1rho} The dependence of the energy density components 
of the universe (in units of $\rho_{0}$)
on the scale parameter $a$ for the parameter values $B=0.6$,
$\Omega_{\Lambda}^{0}=0.7$, and $w=0$.
For large values of the scale factor, $\rho_{m}$ tends to zero, whereas
$\rho_{\Lambda}$ approaches $\rho$ and tends to a constant nonzero value.
The hierarchy $\rho_{m} < \rho_{\Lambda} < \rho$ is maintained for all values of
the scale factor.}
\end{figure}

\begin{figure}
\centerline{\rotatebox{-90}{\resizebox{0.4\textwidth}{!}
{\includegraphics{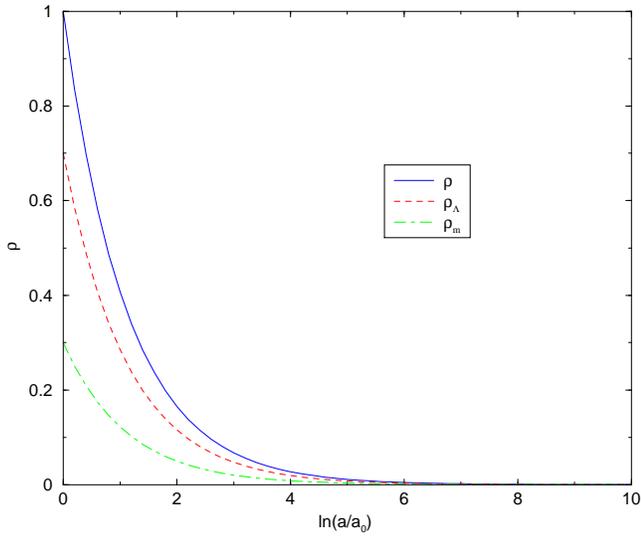}}}}
\caption{\label{fig:2rho} The dependence of the energy density components of the
universe (in units of $\rho_{0}$) for the parameter values 
$B=\Omega_{\Lambda}^{0} = 0.7$ and $w=0$. For large
values of the scale factor $a$, all energy densities approach zero. The hierarchy 
$\rho_{m} < \rho_{\Lambda} < \rho$ is valid for all values of the scale factor.}
\end{figure}

The definition of the cosmological energy scale requires that 
$u_{\rho}^{1/2} \geq 0$. 
This requirement for values $B > \Omega_{\Lambda}^{0}$ leads to the existence
of the {\em maximal value of the scale factor}. In general, the expression for
the largest value of the scale factor $a_{max}$ is 

\begin{equation}
\label{eq:amax}
a_{max} = a_{0} \left( \frac{B- \Omega_{\Lambda}^{0}}{ 1- \Omega_{\Lambda}^{0}}
\right)^{-\frac{2}{3(1-B)(1+w)}} \, ,
\end{equation}
while for the special case $B=1$, we find

\begin{equation}
\label{eq:amaxB1}
a_{max} = a_{0} e^{\frac{2}{3(1-\Omega_{\Lambda}^{0})(1+w)}} \, .
\end{equation}

\begin{figure}
\centerline{\rotatebox{-90}{\resizebox{0.4\textwidth}{!}
{\includegraphics{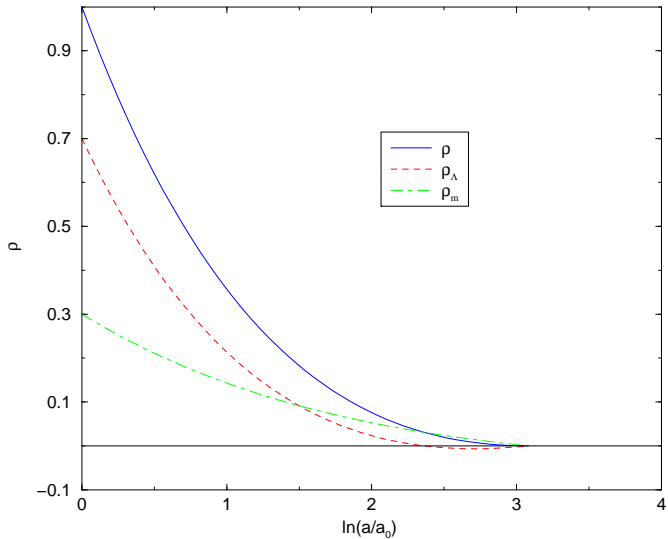}}}}
\caption{\label{fig:3rho} The dependence of the energy density components 
(in units of $\rho_{0}$) of the
universe for the parameter values $B=0.85$, $\Omega_{\Lambda}^{0}=0.7$, 
and $w=0$. The scale factor $a$ has a maximal
value $a_{max}$. For scale-factor values
close to $a_{max}$, the $\rho_{\Lambda}$ becomes negative, while $\rho_{m}$
surpasses $\rho$.}
\end{figure}

\begin{figure}
\centerline{\rotatebox{-90}{\resizebox{0.4\textwidth}{!}
{\includegraphics{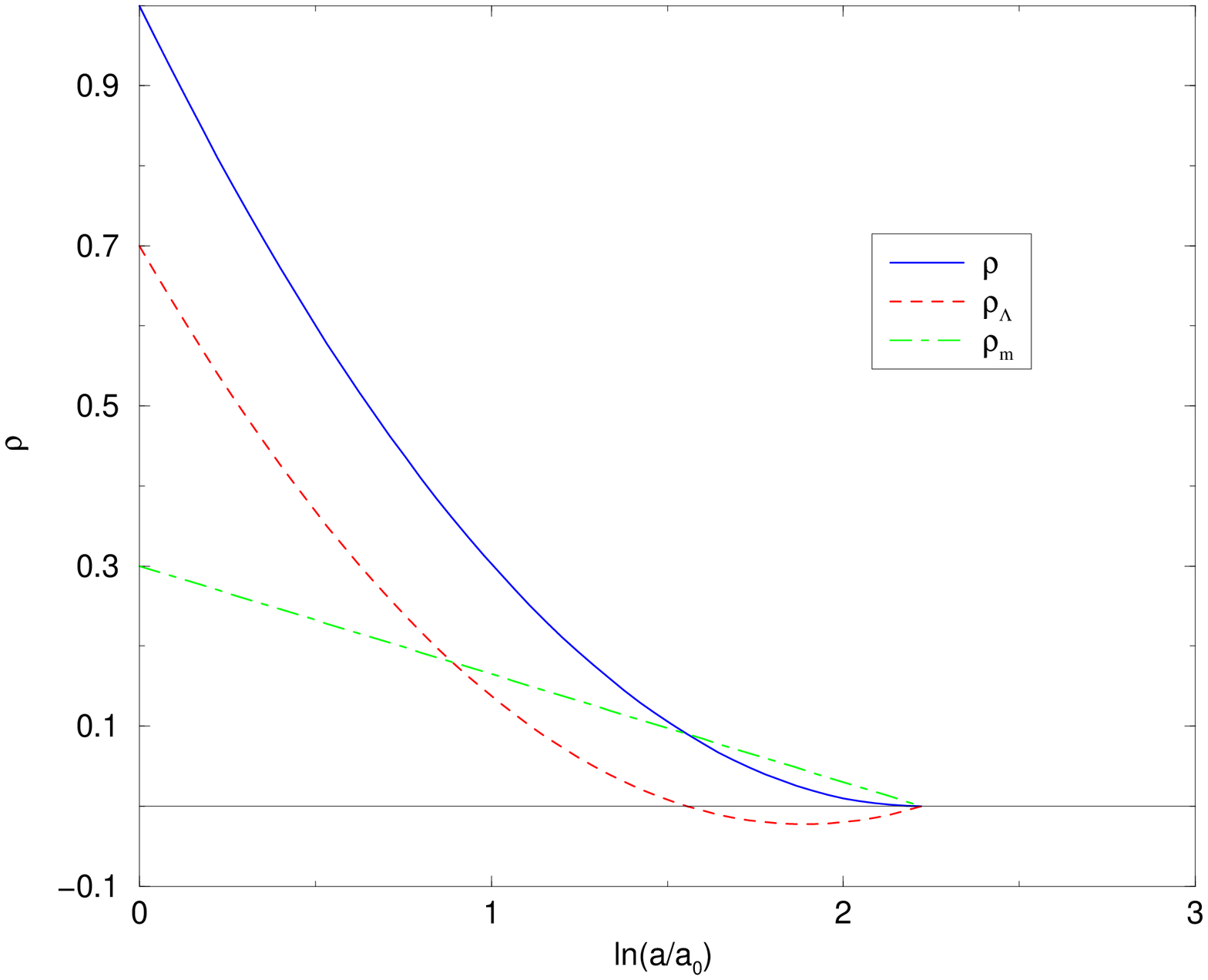}}}}
\caption{\label{fig:4rho} The dependence of the energy density components of the
universe (in units of $\rho_{0}$) 
for the parameter values $B=1$, $\Omega_{\Lambda}^{0}=0.7$, and $w=0$. 
The scale factor $a$ has a maximal value $a_{max}$. For scale-factor values
close to $a_{max}$, the $\rho_{\Lambda}$ becomes negative, while $\rho_{m}$
surpasses $\rho$.}
\end{figure}

\begin{figure}
\centerline{\rotatebox{-90}{\resizebox{0.4\textwidth}{!}
{\includegraphics{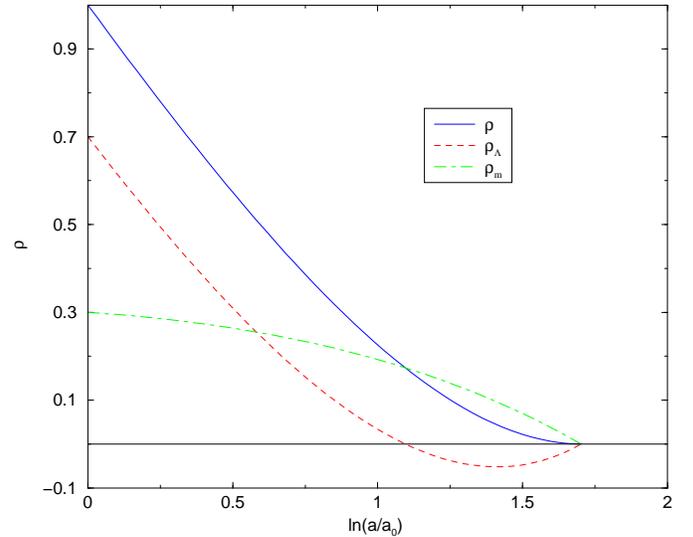}}}}
\caption{\label{fig:5rho} The dependence of the energy density components of the
universe (in units of $\rho_{0}$) for the parameter values $B=1.2$,
$\Omega_{\Lambda}^{0}=0.7$, and $w=0$. The scale factor $a$ has a maximal
value $a_{max}$. For scale-factor values
close to $a_{max}$, the $\rho_{\Lambda}$ becomes negative, while $\rho_{m}$
surpasses $\rho$.}
\end{figure}

Finally, we can incorporate the laws of change of the total energy density
with the scale factor into the Friedmann equation for the flat universe ($k=0$):
 
\begin{equation}
\label{eq:friedmann}
\left(\frac{\dot{a}}{a} \right)^{2} = \frac{8 \pi G}{3} \rho \, .
\end{equation}
Solving the Friedmann equation, we obtain the law of the evolution of 
the scale factor

\begin{equation}
\label{eq:aodt}
a = a_{0} \left[ \frac{1-B}{\Omega_{\Lambda}^{0} - B} e^{\frac{3}{2} H_{0} (t -
t_{0}) (\Omega_{\Lambda}^{0}-B) (1+w)} -
\frac{1-\Omega_{\Lambda}^{0}}{\Omega_{\Lambda}^{0} - B} \right]^{\frac{2}{3
(1- B)(1+w)}} \, .
\end{equation}
\begin{figure}
\centerline{\rotatebox{-90}{\resizebox{0.4\textwidth}{!}
{\includegraphics{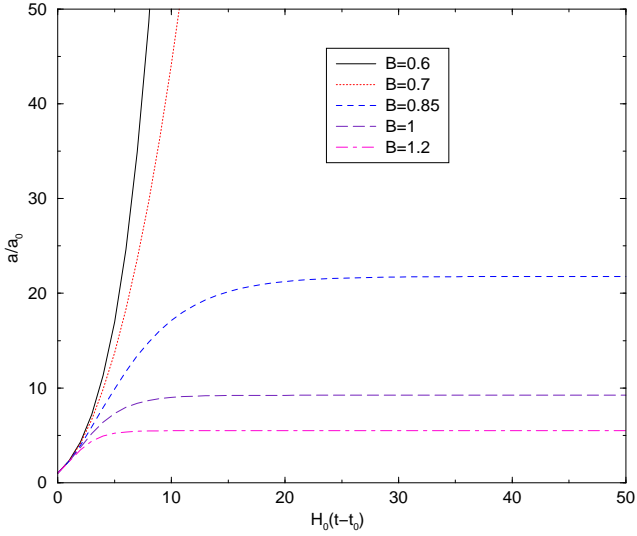}}}}
\caption{\label{fig:scales} Evolution of the scale factor $a$ with time for five
different values of the parameter $B$ and $\Omega_{\Lambda}^{0}=0.7$, $w=0$. 
Notice that for values $B \leq
\Omega_{\Lambda}^{0}$ the universe expands without limits, whereas for 
$B > \Omega_{\Lambda}^{0}$ the scale factor asymptotically tends to $a_{max}$
and reaches it at infinity.}
\end{figure}
For $B=\Omega_{\Lambda}^{0}$, the $a(t)$ law has the form

\begin{equation}
\label{eq:aodtblam}
a = a_{0} \left[ 1 + \frac{3}{2} H_{0} (t -t_{0}) (1+w) (1-\Omega_{\Lambda}^{0})
\right]^{\frac{2}{3(1-\Omega_{\Lambda}^{0})(1+w)}} \, ,
\end{equation}
while for $B=1$, we obtain

\begin{equation}
\label{eq:aodtB1}
a = a_{0} e^{\frac{2}{3(1-\Omega_{\Lambda}^{0})(1+w)} \left[ 1 - e^{-\frac{3}{2}
H_{0} (1- \Omega_{\Lambda}^{0})(1+w)(t-t_{0})} \right]} \, .
\end{equation}
An important feature of the evolution for $B > \Omega_{\Lambda}^{0}$ is that,
although there exists a maximal value of the scale factor, it takes the universe 
infinitely long to reach it. Moreover, for any finite time the total energy
density $\rho$ is nonnegative which makes our definition of the 
cosmological energy
scale generally acceptable (and not only for some finite interval of time
starting now). 

The cosmological model described above also has interesting consequences for the 
event horizon $d_{E}(t_{0}) = a_{0} \int_{t_{0}}^{\infty} \frac{d t'}{a(t')}$.

In the case $B=\Omega_{\Lambda}^{0}$, we can give the expression for the event
horizon in the closed form

\begin{equation}
\label{eq:eventhorizon}
d_{E}(t_{0}) = \frac{2}{2 - 3(1-\Omega_{\Lambda}^{0})(1+w))} \frac{1}{H_{0}} \,
.
\end{equation}

\begin{figure}
\centerline{\rotatebox{-90}{\resizebox{0.4\textwidth}{!}
{\includegraphics{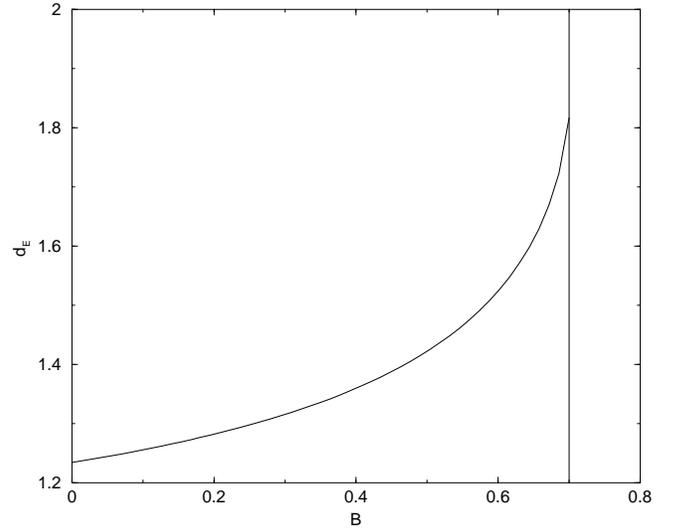}}}}
\caption{\label{fig:hor} The event horizon for the parameter values $B \leq
\Omega_{\Lambda}^{0}$ in units $1/H_{0}$. 
For values $B > \Omega_{\Lambda}^{0}$, the event horizon becomes infinite.
The parameter values used are $\Omega_{\Lambda}^{0}=0.7$ and $w=0$.
The vertical line at $B=\Omega_{\Lambda}^{0}$ 
is introduced as a guidance to the eye.}
\end{figure}
Numerical integration for the values of the parameter $B \leq \Omega_{\Lambda}^{0}$
gives the finite values for the event horizon as depicted in Fig. \ref{fig:hor}.
On the other hand, the event horizon for the parameter values $B >
\Omega_{\Lambda}^{0}$ diverges. This can be easily seen from the fact that in
this parameter range the scale factor is bounded from below and from above:

\begin{equation}
\label{eq:abound}
a_{0} < a(t) < a_{max} \, ,
\end{equation}
which leads to

\begin{equation}
\label{eq:dbound}
a_{0} \int_{t_{0}}^{\infty} \frac{d t'}{a_{0}} > d_{E}(t_{0}) > 
a_{0} \int_{t_{0}}^{\infty} \frac{d t'}{a_{max}} \, .
\end{equation}
As both the upper and the lower bound in the above inequality diverge, 
the event horizon is infinite in this parameter range. 
 
Let us now discuss the main features of our scenario in a more detail. As
shown in our figures \ref{fig:1rho}-\ref{fig:scales}, 
several different outcomes concerning the fate
of the universe as well as the scale factor evolution are possible, depending,
in essence, on how the parameter $B$ is related to $\Omega_{\Lambda }^0 $.
Once $B$ is picked, the parameter $A$ is given by (\ref{eq:A}). 
Our treatment of
$A$ and $B$ as free parameters means that our analysis is, on the whole,
model independent (i.e., we do not restrict ourselves to 
a given particle theory). Let
us mention, for instance, that considering zero-point energies only, we have
$B= \frac{1}{(4 \pi)^2} \frac{37}{4}$ in the standard model (treating neutrinos
as massless). 

First, consider case $B < \Omega_{\Lambda }^0 $, as depicted in fig.
\ref{fig:1rho}. We see
that $\rho_{\Lambda }$ approaches a positive constant 
$\rho_{0} \left( \frac{\Omega_{\Lambda}^{0} - B}{1 - B} \right)^2$ when $a
\rightarrow \infty $, whereas $\rho_m $ approaches zero in the same limit.
This means that the universe once entering the dS regime, stays there for an
indefinitely long time, see fig. \ref{fig:scales}. 
In this case, the running of $\rho_{\Lambda }$ plays
no essential role, since it is the asymptotic behavior of $\rho_{\Lambda }$
and $\rho_m $ that matters in determining the ultimate fate. Of course, the
event horizon is finite, see fig. \ref{fig:hor}. Next, consider the limiting case $B =
\Omega_{\Lambda }^0 $ depicted in fig.\ref{fig:2rho}. 
Now both $\rho_{\Lambda }$ and $\rho_m $ tend to
zero when $a \rightarrow \infty $. The hierarchy $\rho_m < \rho_{\Lambda }$
preserved in the whole running is crucial for the ultimate fate, which turns
out to be the same as in the preceeding case, see fig. \ref{fig:scales}. 
The event horizon is
also finite, see fig.\ref{fig:hor}. Now we consider the most interesting case $B >
\Omega_{\Lambda }^0 $ as depicted in figs.\ref{fig:3rho}-\ref{fig:5rho}. 
In all these cases, the
universe left the dS phase relatively soon, when $a \sim a_{0} e^{0.5 - 1.5}$
(depending on $B$), then starting to decelerate. The reason for this is
the change in hierarchy from $\rho_{\Lambda } > \rho_m $ to $\rho_m >
\rho_{\Lambda }$. Soon afterwards, when $a \sim a_{0} e^{1 - 2.5}$ 
(depending on
$B$), $\rho_{\Lambda }$ becomes negative, staying such forever. We recall
that for the ``true'' $\Lambda $ ($\Lambda < 0 $), the universe recollapses
unavoidably. The reason why we have a different fate here is that the
established hierarchy, $\rho_m > \mid \rho_{\Lambda } \mid $, is preserved
also asymptotically, where both components tend to zero energy density.
Stated differently, since
$\mid \rho_{\Lambda } \mid $ goes faster to zero than $\rho_m $ when $a
\rightarrow a_{max}$, the former component never has a chance
to start dominating, a feature crucial for having recollapse within a finite
time interval. Finally, we need to explain why $a_{max} < \infty $ in this
case, see fig. \ref{fig:scales}. Stating formally, the recollapse here occurs at infinity
(i.e., for $t = \infty $), when $\mid \rho_{\Lambda } \mid $      
finally reaches $\rho_m $. Since any recollapse (starting whenever) requires
$a_{max} < \infty $, we have $a(t = \infty ) = a_{max} < \infty $, as seen
from  (\ref{eq:amax},\ref{eq:amaxB1}) and fig. \ref{fig:scales}. 
The matter dominance in the asymptotic regime, helped with
the influence of negative $\Lambda $, makes the event horizon diverge, thus
alleviating the problems with strings, see fig. \ref{fig:hor}.

The results on the evolution of the universe presented in this paper refer to 
the case of a flat universe (k=0). For this case, it is
possible to obtain analytical expressions for the time evolution of the scale
factor $a(t)$. In the cases of open (k=-1) or closed (k=+1) universes, the
curvature effects may become important in the considerations of the asymptotic
evolution of the scale factor. In these cases, however, it is not possible to
maintain the same level of analytical tractability.
 
Finally, one would like to know if the above scenario is capable of solving
the well-known problems that one encounters when explaining dark energy with a
``true'' CC. The first problem is to ensure a phase of inflation, an epoch
when vacuum energy dominated other forms of energy density, characterized
by $\rho_{vac} \gg \rho_{\Lambda}^0$. Let us also mention the so-called
coincidence problem, which requires an extremely large hierarchy between
$\rho_R$ and $\rho_{\Lambda}$ at the epoch just after the inflation
(contrary to the expectation from equipartition), in order to correctly
reproduce $\rho_{\Lambda}$ today, with $\Lambda = constant$. Since both of
the  problems are addressed by ``quintessence'', one may wish to know
whether the above scenario could mimic some of quintessence models. This
amounts to study the running (\ref{eq:runneq}) backward into a distant past. 
Whereas at 
extremely low energies we could keep only the first two terms in the
expansion of (\ref{eq:lamatmu}) and perform the analysis without referring to any particular
model, this is no longer possible at high energies. Now we need all
peculiarities of the underlying particle theory: particle masses, all
relevant interactions, symmetry breaking properties, etc. 
For instance, the description of the running of the cosmological constant 
energy density at high scales $\mu$ requires the knowledge of the number and 
individual masses of the heaviest degrees of freedom (and not just a small number
of specific combinations of these parameters like coefficients $A$ and $B$). 
Such complete information is currently not available even for the simplest
theories, like the standard model. There is an additional technical complication
in the region of high scales. Namely, one has to take into account the
contributions coming from mass scales differing many orders of magnitude, which
hinders a straightforward numerical treatment.
In addition,
higher-loop contributions together with the running of masses and couplings,
all of which we have obviously disregarded in (\ref{eq:runneq}), 
may no longer be irrelevant
at high energies. We feel, though, that it would certainly be of
importance to address the coincidence problem even in a toy model, in view
of the obvious connection between $\rho_{\Lambda}$ and $\rho_m$ as expressed
in (\ref{eq:lambdarho}). Since our examples show a clear tendency for $\rho_{\Lambda}$ to
grow faster than $\rho_m$ for $t < t_0$, one may object that there is no
room in our scenario to explain the discovery of the supernova SN 1997ff, 
which is believed to signal a switch from cosmological acceleration
to deceleration \cite{14}. 
Once again we must stress that this potential drawback
could result from our ignorance of the full theory. We recall \cite{BGHS} that
the smallness of the parameter $A$ requires a fine-tuned relation among
particle masses of the underlying theory. One may therefore imagine that the
running of masses, although believed to be weak at such energies and
therefore ignored in (1), could easily change the sign of $A$, thereby depleting
$\rho_{\Lambda}$ at higher energies.

As a last step, it would be very useful to compare the 
effects originating from
the matter sector (treated in this paper) with quantum gravity effects on
the running of the cosmological constant. Firstly, we could interpret the
assumption $\rho_{\Lambda}(0) = 0$ as an 
effective inclusion of the effects of quantum 
gravity in our model since within our model there is no mechanism or principle
that would dictate vanishing of $\rho_{\Lambda}$ at $\mu=0$. This choice
certainly represents a rather crude approximation of the effects of quantum
gravity, but, in our view, at the same time singles out the most important of
these effects. Secondly, in the absence of complete understanding of quantum
gravity effects, especially in the infrared region of the scale $\mu$, it is not
possible to say which type of effect (matter sector or quantum gravity) would 
dominate the running of the cosmological constant energy density at 
infrared scales.
Still, our findings of significant (and possibly sign-changing) running of the
cosmological constant energy density coming from the matter sector make
reasonable the assumption
that quantum gravity effects (apart from those setting 
$\rho_{\Lambda}(0)= 0$) do not significantly change the running of the CC 
and the respective consequences on the evolution of the universe.
We see our model as complementary to quantum gravity approaches, especially to
those which draw conclusions from the pure gravity theories assuming that the
matter sector does not alter the pattern from the pure gravity theory 
\cite{Bon}. Our results show that one cannot a priori expect the effects of
the matter sector to play only a secondary role. 
We expect that only a full solution of quantum effects in gravity along with
a careful treatment of contributions coming from the matter sector will
resolve the problem of the running of the cosmological constant unambiguously.

In conclusion, we have studied the renormalization-group running of the
cosmological constant at scales well below the lowest mass in a generic
particle theory. Having specified the renormalization-group scale in
cosmological settings, we considered the ultimate fate of the universe in
such a scenario. As its main feature, our scenario naturally provides 
the possibility that
the universe endowed with a positive cosmological constant nowadays,
necessarily exits a de Sitter regime a few Hubble times afterwards. More
surprisingly, although such a universe exhibits the maximium value of the
scale factor, it never recollapses. Although the running into past times
cannot be provided in a model-independent manner, the relation between
$\rho_{\Lambda}$ and $\rho_{m}$ implied in (\ref{eq:lambdarho}) indicates that,
treated in a specific particle physics model, our approach might also 
contribute to the resolution of the coincidence problem.


\begin{acknowledgments}
This work was supported by the Ministry of Science and Technology of the
Republic of Croatia under the contracts No. 0098002 and 0098011.
\end{acknowledgments}

\end{document}